\documentclass[serv, nonblindrev]{format-informs-serv} 
\MANUSCRIPTNO{}

\OneAndAHalfSpacedXII

\usepackage{xcolor}
\usepackage{soul}
\usepackage{booktabs}
\usepackage{tabularx}
\usepackage{multirow}
\usepackage{threeparttable}
\usepackage[breaklinks]{hyperref}
\makeatletter
    \g@addto@macro\UrlBreaks{\do\-}
\makeatother

\usepackage{natbib}
 \bibpunct[, ]{(}{)}{,}{a}{}{,}%
 \def\bibfont{\small}%
\renewcommand{\theARTICLETOP}{} 

\begin{document}

\RUNTITLE{COVID-19 Vaccine Hesitancy}
\TITLE{An Analysis of COVID-19 Vaccine Hesitancy in the U.S. at the County Level}

\ARTICLEAUTHORS{
\AUTHOR{Hieu Bui$^1$, Sandra Eksioglu$^1$, Ruben Proano$^2$, Sarah Nurre Pinkley$^1$}
\AFF{$^1$Department of Industrial Engineering, University of Arkansas, Fayetteville, AR 72701, \EMAIL{}}
\AFF{$^2$Department of Industrial and Systems Engineering, Rochester Institute of Technology, Rochester, NY 14623, \EMAIL{}}
}
\RUNAUTHOR{Bui et al.}

\defcitealias{cdc_2021_state}{CDC 2021a}
\defcitealias{cdc_2021_county}{CDC 2021b}
\defcitealias{cdc_influenza}{CDC 2021c}
\defcitealias{cdc_data_updates}{CDC 2022a}
\defcitealias{CDC_olderAdults}{CDC 2022b}
\defcitealias{who_2014}{WHO 2014}
\defcitealias{who_2019}{WHO 2019}
\defcitealias{kff_monitor}{KFF 2021a}
\defcitealias{kff_political}{KFF 2021b}
\defcitealias{aspe_vh_method}{ASPE 2021a}
\defcitealias{aspe_vh}{ASPE 2021b}
\defcitealias{voting_results}{MIT 2021}
\defcitealias{dataset_internet}{USCB 2021}
\defcitealias{dataset_poverty}{USCB 2019}
\defcitealias{labor_stats}{BLS 2021}
\defcitealias{fcc_location}{FCC 2021}

\ABSTRACT{
Reluctance or refusal to get vaccinated, referred to as vaccine hesitancy (VH), has hindered the efforts of COVID-19 vaccination campaigns. It is important to understand what factors impact VH behavior. This information can help design public health interventions that could potentially increase vaccine uptake. We develop a random forest (RF) classification model that uses a wide variety of data to determine what factors affected VH at the county level during 2021. We consider static factors (such as, gender, race, political affiliation, etc.) and dynamic factors (such as, Google searches, social media postings, Stringency Index, etc.). Our model found political affiliation and the number of Google searches to be the most relevant factors in determining VH behavior. The RF classification model grouped counties of the U.S. into 5 clusters. VH is lowest in cluster 1 and highest in cluster 5. Most of the people who live in cluster 1 are democrat, are more internet-inquisitive (are more prone to seek information from multiple sources in the internet), have the longest life expectancy, have a college degree, have the highest income per capita, live in metropolitan areas. Most people who live in cluster 5 are republicans, are the least internet-inquisitive, have the shortest life expectancy, do not have a college degree, have the lowest income per capita, live in non-metropolitan areas. Our model found that counties in cluster 1 were most responsive to vaccination-related policies and COVID-19 restrictions. These strategies did not have an impact on the VH of counties in cluster 5.}

\maketitle

\let\clearpage\relax

\section{Introduction}
\label{sec:intro}

Millions of people around the world are still affected by the COVID-19 pandemic. COVID-19 vaccines have proven to be effective in reducing the risk of hospitalizations and death, especially among older adults. Unfortunately, reluctance or refusal to get vaccinated, referred to as vaccine hesitancy (VH), has hindered the goal of the vaccination campaign \citepalias{who_2014}. In 2019, the World Health Organization (WHO) identified VH as one of the ten threats to global health \citepalias{who_2019} because VH impacts the spread of the disease and the number of casualties.  According to the Centers for Disease Control and Prevention (CDC), on January 1, 2022, $62.8 \%$ of the total population in the U.S. were fully vaccinated, ranging from Idaho at $46.3 \%$ to Vermont at $77.5\%$ \citepalias{cdc_2021_state}. Failure to address VH could lead to the emergence of new COVID-19 variants, which would prolong the pandemic \citep{fox_2021}. Additionally, VH impacts the ability of the government and health officials to make accurate demand forecasts for vaccines, leading to vaccine stock-out and wastage.


Surveys, polls, and questionnaires are the preferred methods to evaluate attitude toward vaccination. Although these methods provide good estimates of VH attitude, and provide insights into why people may be hesitant to get vaccinated, they are expensive and time-consuming. Additionally, these methods are limited in scope because they present VH periodically, at a single point in time. Other limitations arise because of the method used for data collection, the targeted population, and the set of questions asked \citep{khubchandani_2021}. The literature points to discrepancies in the definition and the context of VH, which could mislead survey takers \citep{eve_2013, kumar_2016}. Nevertheless, the CDC and  Department of Health in the several U.S. States have collected and made available to the public data about vaccine uptake at the county level. This data can be used to develop measures of VH to enable decision makers to evaluate changes in its behavior over time and compare it across different population groups. In our proposed research \emph{we develop a metric of VH behavior using several sources of publicly available data}.  We compare its performance to VH estimates developed from surveys. 
Several studies focus on determining what factors impact VH attitude and behavior. Some of these studies analyze the role of social media on people's attitude towards vaccination, and others use population demographic, social and economic factors to explain why people are not vaccinated. In our study, we group the different factors which impact VH into static and dynamic factors. Static factors, such as gender, race, ethnicity, political affiliation, etc., do not change frequently and, therefore, cannot explain dynamic changes of VH in a community over relatively short periods of time, such as a week, a month, or a year. Static factors allow establishing a baseline to explain how likely individuals are to be affected by dynamic factors. Dynamic factors change over time and can signal the community response to federal and local policies, community interventions, and comments from public policy influencers. For instance, the Internet and social media are increasingly used to share real-time opinions about health topics, including COVID-19 vaccines \citep{andrew_2020, hoffman_2019}. Users can be exposed to misinformation and negative comments contributing to VH \citep{garett_2021}. It is of interest to understand the impact that the dynamic factors have on amplifying the effect of static factors on VH rates.
A dearth of research on VH points to a knowledge gap that limits the ability to study the impact of different factors on the changes observed in the estimation of VH attitude and behavior over time \citep{fridman_2021, king_2021, kff_monitor}. This observation motivated the following research question, which we study: \emph{What are the main factors that impact VH behavior? Under what conditions are some factors more predominant than others?}  

Sufficient and high-quality data related to COVID-19 is available at high levels of aggregation, such as, at the state and national levels. However, we notice a lack of data at the county and zip code levels. Much of the available data (such as, data from Twitter, or vaccination records) comes from large urban areas. For example, California and Virginia do not report to CDC vaccination records for counties with less than 20,000 and 10,000 population, respectively. Thus, national level projections of VH are dominated by the large volume of data collected in highly populated urban areas. Lack of data often presents missed opportunities to explore VH further. This observation motivates the following research question which we study:  \emph{In what  meaningful clusters should counties be aggregated to support efforts of overcoming VH?} Our findings could help develop strategies to address specific challenges leading to VH in these clusters.

With guidance from the CDC, Federal and States public health authorities sought effective strategies to increase vaccination uptake in the U.S.  Some of these strategies are school closing, workplace closing, canceling public events, restrictions on international travel, public information campaigns, etc.  Many states also introduced financial incentives, ranging from small rewords, such as, a free beverage, or a gift card to lotteries that give vaccinated individuals a chance to win large prizes \citep{thirumurthy_2022}. It is of interest to evaluate how effective these strategies were in reducing VH. This observation motivated the following research question, which we study:  \emph{How did vaccination-related policies, interventions, and COVID-19 restrictions impact VH in the U.S.? Were these restrictions as effective in different counties within the U.S.?} 

We present a systematic, data-driven framework to help us understand VH and provide answers to the aforementioned research questions. This framework includes a machine learning (ML) algorithm that uses data from various sources to determine what factors impact VH. We use the Goodness of Variance Fit method to determine 5 clusters, and use
the RF classification model to group counties of the U.S into these clusters. The model is used to estimate changes in VH behavior of different clusters over time and space. These estimates can be used to complement the results of surveys. The outcomes of the models are validated using data from surveys conducted by the U.S. Department of Health and Human Services \citepalias{aspe_vh_method} and the Delphi project \citep{salomone_2021}. 

The remainder of the paper is organized as follows.  Section \ref{lit} provides a summary of the existing literature. Section \ref{sec:method} summarizes the modeling framework proposed and model validation. Section \ref{sec:discussion} presents a discussion of the results. Finally, Section \ref{sec:conclusions} provides a summary and conclusions of the proposed study. 
\section{Literature Review}
\label{lit}
Several studies focus on determining the factors that drive VH. Some of the studies evaluate a number of putative predictors of vaccination willingness and hesitancy. Depending on the field of study, these factors range from psychological to sociological and economical. The WHO’s Strategic Advisory Group of Experts on Immunization (SAGE) presents a concise `3Cs' model to understand vaccination behavior: \emph{confidence}, \emph{convenience}, and \emph{complacency}. \emph{Confidence} is defined as trust in the vaccines' effectiveness and safety, the competence of health services, and policymakers' motivations. \emph{Convenience} refers to the accessibility to the vaccines, related services, and the willingness to pay for the vaccines. Vaccine \emph{complacency} refers to the perceived risk of contracting the disease, and the perceived impact that the disease can have on one's life \citepalias{who_2014}. This model has been extended to incorporate factors such as, the \emph{collective} responsibility and willingness  to protect others by getting vaccinated \citep{betsch_2018}; the impact of \emph{communication} and \emph{context} due to (mis)information from social media platforms \citep{razai_2021}; etc. Our proposed model evaluates the impact of factors related to \emph{confidence}, \emph{convenience}, \emph{complacency}, and \emph{communication}. However, we do not exhaustively explore these factors in our research.

Traditional methods to evaluate VH, conduct empirical studies using data collected via surveys. Several surveys were conducted to capture the intention, readiness, and willingness to get a COVID-19 vaccine. Some of these surveys were deployed before, and others after the COVID-19 vaccines were approved by the Food and Drug Administration (FDA). A survey of 5,009 American adults,  conducted in May 2020, indicated that 31.1\% of the respondents did not intend to get vaccinated due to concerns about vaccine safety and effectiveness \citep{callaghan_2020}. Another 2020 survey of health care workers revealed vaccine effectiveness and safety as the primary reasons for hesitating to get vaccinated \citep{meyer_2021}. A poll by the Kaiser Family Foundation (KFF) revealed that 62\% of the participants were concerned about vaccine effectiveness and safety. These participants believe that social-political pressures due to the 2020 presidential elections in the USA, led to a rushed approval of the COVID-19 vaccines \citepalias{kff_political}. Another national-level assessment of VH via a community-based sample of adult population revealed that individuals who had low education, low income, or perceived threat of getting infected to be high, were more likely not to get COVID-19 vaccine \citep{khubchandani_2021}. This study also found VH to be higher among African-Americans (34\%), Hispanics (29\%), those who had children at home (25\%), rural dwellers (29\%), people in the northeastern US (25\%), and those who identified as Republicans (29\%). Several studies determine that trust on COVID-19 vaccine \citep{wang_2022}, healthcare workers, healthcare system,  science, and policymakers who design vaccination strategies, are important factors in reducing VH. This mistrust is due to misinformation and rumors. Strategies to build trust are improving vaccine literacy, clarifying misinformation and rumors, and providing verified information.    

Additional studies related to VH were conducted after vaccines were approved by the FDA. For example, the New York Times (NYT) used surveys and vaccine administration data to analyze VH at the county level. It was found that both, willingness to get vaccinated (VH attitude), and the actual vaccination rate (VH behavior) were on the average lower in counties where most residents voted for Republicans during the 2020 presidential elections \citep{nyt_political}.  
Several other efforts were carried out to monitor the VH on a large scale, over time. The  U.S. Department of Health and Human Services via the U.S. Assistant Secretary for Planning and Evaluation (ASPE) developed a method to predict VH rates using Household Pulse Survey (HPS) data. ASPE captured VH by analyzing the responses from a survey's question regarding the intention to get  vaccinated  \citepalias{aspe_vh_method}. A research group, in collaboration with Facebook, used a survey tool to monitor the spread of COVID-19. Facebook users were randomly selected and and asked about vaccination intent \citep{salomone_2021}. This tool allows measuring VH across different geographic and demographic groups in the U.S.   

Using surveys is advantageous to understand why people hesitate to get vaccinated. However, they inherit some disadvantages. For example, modifying a survey's questions makes it difficult to compare survey results from different time periods in order to determine trends in VH.  Additionally, surveys are expensive and time consuming to administer and unless they are administered continuously, they offer an static picture of VH. To overcome these challenges, we propose a metric of VH behavior that is calculated using data about vaccination uptake.

Several sources  track COVID-19 related data, such as, the number of people vaccinated, the number of hospitalization, the number of deaths, social media postings and news articles, etc. The vast amount of data available  has attracted the attention of researchers developing natural language processing (NLP) and machine learning (ML) algorithms to study VH. ML techniques and statistical analysis tools have been used to study infectious diseases such as, Measles \citep{carrieri_2021}, Human Papillomavirus (HPV) \citep{du_2020}, etc. \citep{carrieri_2021} propose a random forest classifier to predict VH  of pediatric vaccines. They found employment rate and recycling efforts to be the two most relevant factors to determine VH of a municipality. \citep{chandir_2018} propose several ML algorithms (i.e., random forest, recursive partitioning, support vector machines, and C-Forest) to predict the likelihood of a child defaulting from subsequent immunization. \citep{bell_2019} develop a LASSO logistic regression model to identify children who are at risk of not being vaccinated against MMR. \citep{lange_2022} use an ordinary least square regression analysis and a random forest algorithm to evaluate the impact of race, poverty, age and political affiliation on COVID-19 vaccination rates. It was determined that counties with higher percentage of Republicans, higher proportion of African Americans, and  higher poverty rate had lower vaccination rates.

Most recently, we have seen an increased interest in studies that use supervised and unsupervised ML algorithms to mine data from social media outlets, such as Twitter and Facebook, in order to help us understand the impact of public discussions on health-related issues and behaviors, such as, VH. Many people still hold negative sentiments about vaccines due to misinformation, lack of trust, and worry about side effects   \citep{ali_2021}. These sentiments are often revealed via posts in social media.
Work by  \citep{deiner_2019} analyzes about 58K Facebook posts and 83K tweets from 2009 to 2016 to study the attitude towards the measles vaccine. 
\citep{wilson_2020} show a statistically significant relationship between disinformation campaigns on social media and the VH for pediatric vaccines. \citep{to_2021} use multiple ML algorithms to analyze 1.5M  tweets in order to determine anti-vaccination contents in this (Twitter) social media platform. \citep{yousef_2021} perform a sentiment analysis on 4.5M tweets collected during January 2020 to January 2021. The study concludes that the number of negative discussions about COVID-19 vaccines was higher than those favoring vaccines. The intensity of these discussions vary across countries. \citep{chandrasekaran_2020} used 13.9M tweets to analyze VH sentiments. They  determined a total 26 topics,  ranging from the source of the pandemic to the government response, and measured people sentiment in each topic.  

Similar to this literature,  we propose a ML model that uses publicly available data from surveys, social media, the Internet, etc. to understand VH behavior. However, different from the literature, the proposed model uses static and dynamic features at the county level.  Such a model, by evaluating the impact of  economical, social and political factors and public opinion on VH at the county level, can help public health authorities in developing tailored strategies focused on increasing the uptake of COVID-19 vaccines. 

\subsection{Research Contribution}

The following is a list of major contributions of the proposed research. \par ($i$) The proposed \emph{ML model determines what factors impact the changes observed in the COVID-19 VH behavior at the county level over time}. This model considers several static and dynamic features, such as, political affiliation, google search insights, Stringency Index, education level, etc. The model uses these factors to cluster counties together. We provide a through description of each cluster and discuss the impact of these features on the corresponding VH.\par  

($ii$) The model \emph{uses a large amount of data collected from several public sources during a period of 9 months, January to October of 2021}. Several important observations are made, which we discuss in Section \ref{sec:discussion}. \par  ($iii$) We develop \emph{a measure of VH behavior} which can be monitored over time. This measure enables the study of VH  using publicly available data and without relying on the use of surveys. The viability of this metric is evaluated via a comparisons with data related to VH obtained from two major surveys, one conducted by the ASPE \citepalias{aspe_vh_method}, and the other by the Delphi project \citep{salomone_2021}.

\section{Method}
\label{sec:method}
\begin{figure}[t]
\FIGURE
{\includegraphics[width=0.95\textwidth]{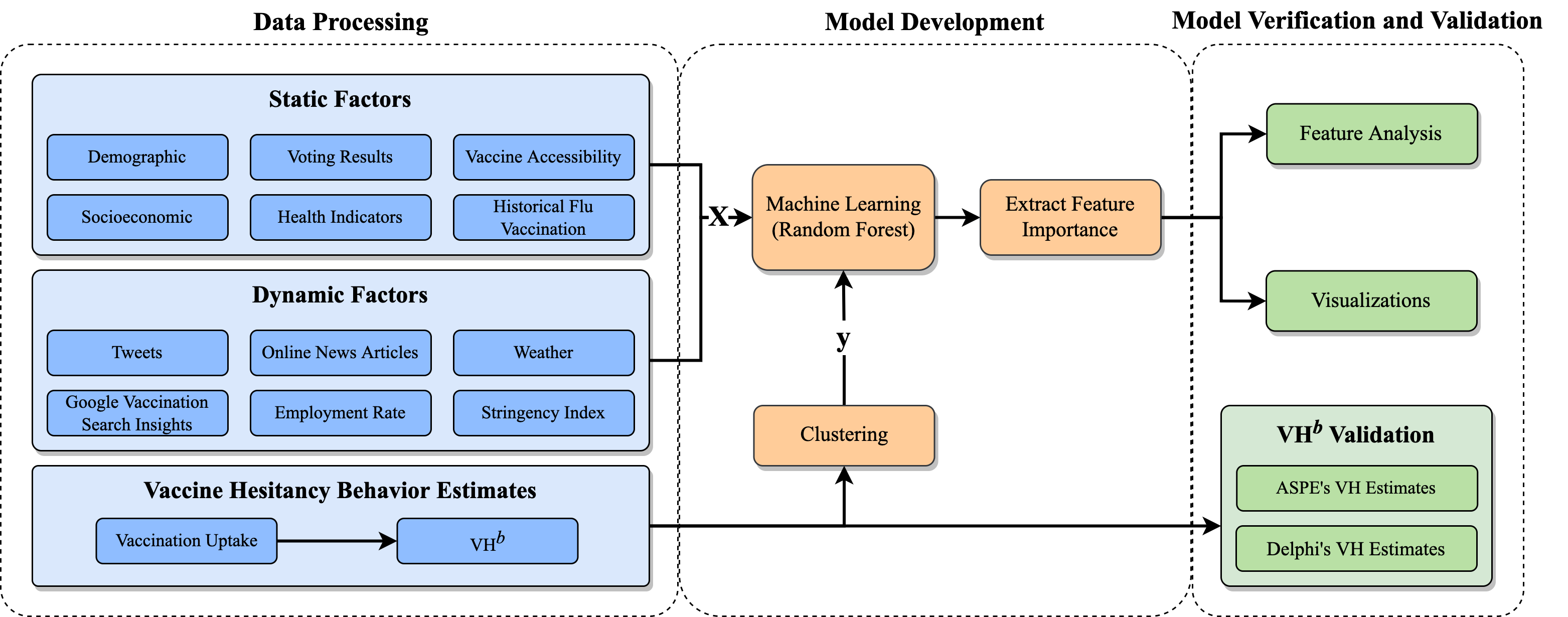}}
{Flowchart of the proposed research.\label{fig:framework}}
{}
\end{figure}

An overview of the proposed modeling framework is illustrated in Figure \ref{fig:framework}. This framework consists of three major parts, which are data acquisition and processing, model development, and model validation. The data processing focuses on evaluating the data to determine and handle inconsistencies. Model development focuses on a Random Forest (RF) classification model, and extraction of feature importance.  Lastly, the outcomes of the proposed model are validated and verified. 

\subsection{Data Acquisition and Data Processing}

\label{sec:dataAcquisition}
{\bf Data Acquisition:} We collected county-level data for the period January 25, 2021, to October 31, 2021, using multiple open-access datasets. We only used the data for which we could find the corresponding county specific Federal Information Processing Standards (FIPS) code. This code is important for fusing together different sources of data to support our model development. As a result, we were able to collect (and use) county-level data for 48 contiguous states and the District of Columbia.  

Our primary source of data is the ``COVID-19 Open-Data,'' which is published in the Google Cloud Platform and GitHub \citep{wahltinez_2020}. This dataset contains demographic characteristics, health indicators, vaccination access, weather conditions, and COVID-19 search trends. We obtained  estimates of VH from ASPE. ASPE utilized the data collected via the HPS to estimate the VH at the county and state levels. This data is available at the county-level via CDC's data repository \citepalias{aspe_vh}. It contains information for a total of 3,142 counties. We  collected data about vaccine uptake from CDC's data repository  \citepalias{cdc_2021_county}. The influenza vaccination coverage for  2018-2019 and 2019-2020 seasons was extracted from the same repository \citepalias{cdc_influenza}. U.S. Census Bureau provides the proportion of households that have internet subscriptions in each county \citepalias{dataset_internet}, and the corresponding poverty status \citepalias{dataset_poverty}. The U.S. Bureau of Labor Statistics provides data related to the labor force and unemployment rates  \citepalias{labor_stats}. The county-level 2020 presidential voting results were extracted from Harvard Dataverse \citepalias{voting_results}.

We collected Twitter postings (tweets) for the period of study. We developed a custom tweet scraper using Twitter's application programming interface (API).  Retweets are excluded from the dataset. Additionally, the custom scraper only searches for tweets that have geographical metadata. 

The GDELT Project introduced a database of news articles, available online, related to COVID-19 vaccinations \citep{gdelt_blog}. This database is accessible through the Google BigQuery data warehouse. We used this database to create a dataset that contains articles, news published online, and the corresponding average tone for every county considered in this study \citep{bigquery_gdelt}. We  only considered  articles from U.S. sources that mentioned COVID-19 vaccines. 

{\bf Data Processing:} Two of the datasets we used require special attention as they contain a large number of fields, such as, weather-related data and Google search trends. For example, weather-related data contains fields, such as, the minimum and maximum temperatures, precipitation amount, wind speed, etc. Some of these fields are highly correlated, such as, the maximum and the minimum temperatures of a given day. We reduced the complexity of these datasets by using the principal component analysis (PCA) method. The goal of PCA is to reduce the number of fields while maintaining valuable information from the dataset \citep{abdi_2010}. PCA provides new fields that are linear functions of the fields from the original dataset. These fields are called principal components (PCs). Weather-related dataset contains 7 fields and the Google search trends contains 22 fields. Our PCA provided 3 PCs of weather-related data, and 5 PCs of Google search trends. These PCs  provide an explained variance of at least 95\%. 

GDELT contains online news articles related to COVID-19 vaccinations. The dataset provides an article-level sentiment attribute that ranges from -100 to 100. A 100 corresponds to an extremely positive tone, a -100 corresponds to an extremely negative tone, and a 0 corresponds to a neutral tone. These values are determined based on a count of the words that have a positive/negative emotional connotation in the article. Each record of the GDELT dataset corresponds to a location mentioned in each news article. Thus, the dataset contains duplicate entries of the same news article, especially in sites that have continuously tracked events since the onset of the pandemic. Hence, we process this data to eliminate duplicates. The processed dataset contains 1,059,758 online articles. We use this data to calculate an average tone per county. Counties that are not mentioned in the news, have a neutral tone of zero.
 
The Twitter data is processed to ensure its compatibility with the rest of the data collected. We noticed that there is a location associated with each tweet. This location could either be the location of the tweet or the location of the account holder. Since we are interested about county-level data, we developed a process to facilitate data collection. We used the FCC Census Block Conversion API, which allowed us to use the longitude and latitude coordinates of a tweet to determine the county it belongs to \citepalias{fcc_location}. For tweets that did not have a location, we used the Twitter Places lookup and alias name lookup  to determine the location of the person who tweeted \citep{grammakov_2020}. As a result, we gathered 588,686 COVID-19 related tweets for which we know the county of origin.

We further processed tweets to remove extra spaces, hyperlinks, and tweet-specific syntax (such as, user mentions of the form ``@username'' and hashtags of the form ``\#hashtag''). To assess the sentiment of a tweet, we used the Valence Aware Dictionary and sEntiment Reasoner (VADER) \citep{hutto_2014}. Because VADER was developed with social media text in consideration, it can handle sentences with slang, emoticons, emojis, and punctuation. Thus, no further steps were required for preparing the inputs for the sentiment analysis step. Next, we continued to process each tweet for topic modeling by forming $n$-grams (i.e., sequence of $n$ words that frequently appear together), filtering out stop words and punctuation, removing slang, lowering texts, performing text tokenization and lemmatization. Lemmatization is important to reduce redundancy in the text (e.g., converting ``studies'' and ``studying'' to ``study''). Some noticeable bi-grams and tri-grams included in our dataset are \emph{side\_effects}, \emph{tested\_positive}, \emph{relief\_bill}, \emph{social\_distancing}, \emph{shut\_down}, \emph{joe\_biden}, \emph{biden\_administration}, \emph{full\_approval}, \emph{operation\_warp\_speed}, and \emph{emergency\_use\_authorization}.

{\bf Challenges with Data Acquisition:} The CDC uses multiple sources for collecting COVID-19 vaccine uptake data. These sources are jurisdictions, pharmacies, and federal entities. However, aggregating this data to determine trends in vaccination uptake is a challenge because the timing and methods used for data reporting vary by entity. For this reason, the dataset includes a metric called \emph{``completeness''} which represent the proportion of valid records. A valid record has the correct ``county of residence'' information. 

\begin{figure}[t]
\begin{center}
\caption{a) State-based, the average percentage of records with valid FIPS. (b)-(d) The number of fully vaccinated over time in counties in Florida, Virginia, and Texas. Grey lines represent the number of fully vaccinated per county. Crimson lines represent the number of missing ``county of residence'' records.} \label{fig:avg_completeness}
\includegraphics[width=0.9\textwidth]{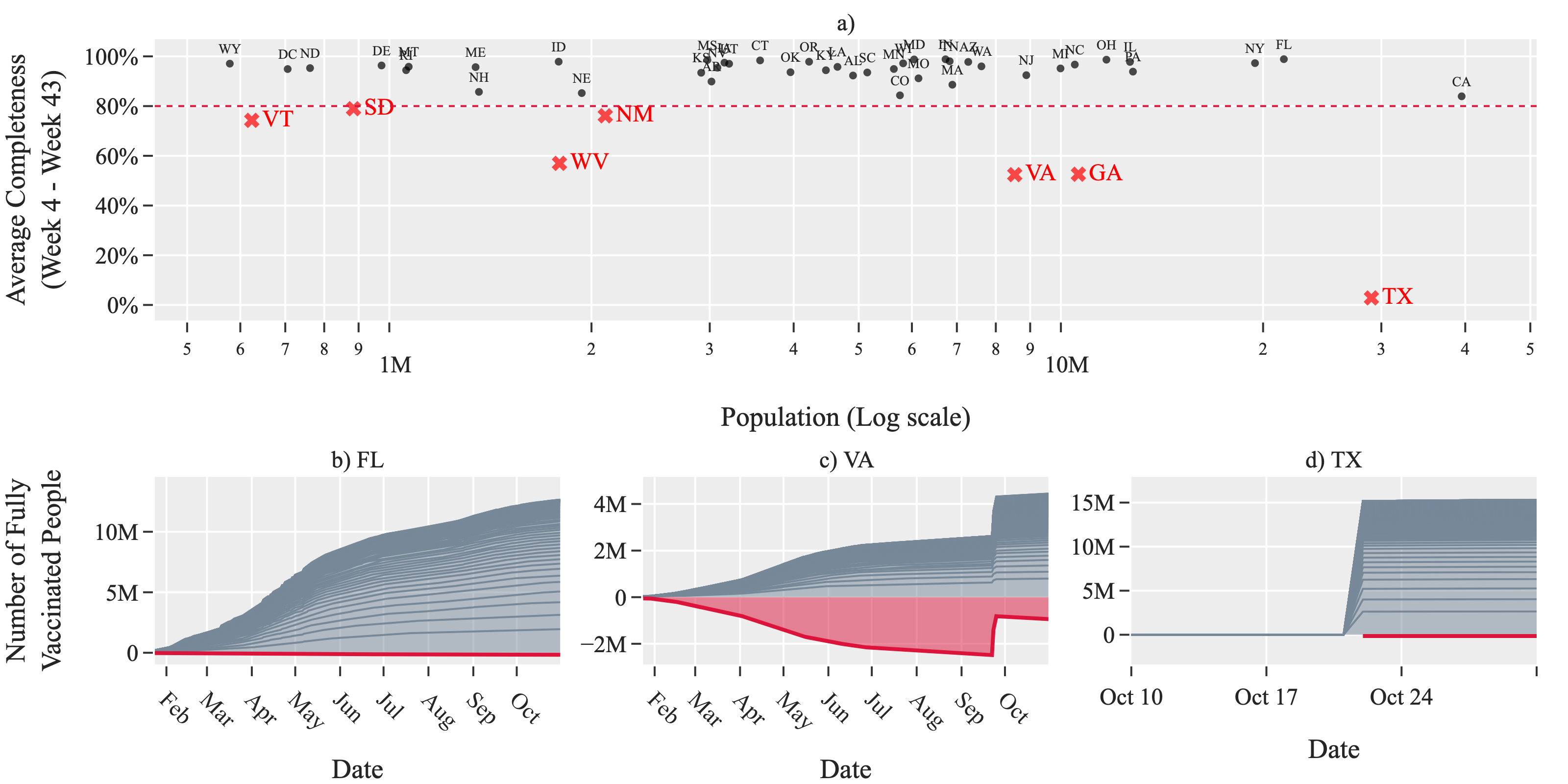}
\end{center}
\end{figure}

Figure \ref{fig:avg_completeness} illustrates the \emph{``completeness''} and the number of fully vaccinated in the U.S. \citepalias{cdc_2021_county}. We noticed that approximately 17\% of the U.S. population lives in 7 states missing, on average, at least 80\% of the information about the ``county of residence" (see Figure \ref{fig:avg_completeness}a). For example, there was no vaccination-related data in all 254 counties of Texas until late October 2021 (see Figure \ref{fig:avg_completeness}d). States with many missing ``county of residence'' records appear to have lower than expected vaccination rates (see Figure \ref{fig:avg_completeness}c). Additionally, states such as, California and Virginia do not report to CDC data for counties with less than 20,000 and 10,000 population, respectively. An update of this dataset on September 24, 2021, reduced the number of missing data for Virginia (as illustrated in Figure \ref{fig:avg_completeness}c). These abrupt changes in the size and accuracy of this dataset impact the accuracy of our estimates of the VH score. 

The aforementioned challenges were the reason why we independently collected vaccination data from the State Health Department websites and \emph{Covid Act Now} API for Texas, Georgia, Virginia, West Virginia, New Mexico, South Dakota, and Vermont. The corresponding results are summarized in Table \ref{tab:vaccinationData}. Finally, for the counties that we could not find data  about vaccination uptake, we substituted the missing value with the average value of vaccination uptake of the neighboring counties.  

\begin{table}
    \TABLE
    {Vaccination related data extracted from CDC and other sources. \label{tab:vaccinationData}}
    {\begin{tabular}{@{}p{0.12\linewidth}@{\quad}p{0.11\linewidth}@{\quad}p{0.17\linewidth}@{\quad}p{0.5\linewidth}@{}}
    \hline
    \up Source & Number of Counties & Total Population Covered & Note\down\\
    \hline\up
    CDC Only  & 2,394/3,143 & 272.5M (83.21\%) &Counties in TX, GA, VA, WV, NM, SD, VT are omitted due to missing ``county of residence'' \\
    Aggregated & 3,095/3,143 & 325.2M (98.12\%) & Some rural counties in SD, TX, VA are excluded \down\\
    \hline
    \end{tabular}}
    {}
\end{table}

\subsection{Model Development}

{\bf An Estimate of Vaccine Hesitancy Behavior:} 
VH is defined as the delay in acceptance or refusal of available vaccines. We noticed the  difference between VH attitude and VH behavior. While surveys, such as, HPS measure the attitudes towards vaccines, vaccine uptake is a  measure vaccination behavior. The data collected by HPS is a single data point that might not help explain changes in behavior and attitude over time, especially at the county level which are due to vaccination related policies and mandates, or due to fear of infection from new variants of COVID-19 virus. Changes in vaccination uptake over time indicate changes of VH behavior. This is the reason why we use the data about vaccine uptake to develop an estimate of VH behavior. 

Let $\delta_{it}$ represent the percentage of unvaccinated population at county $i$ in week $t-\ell$ that was vaccinated during weeks $t-\ell$ to $t$ ($l\geq 1$). We use equation \eqref{eq:roc} to calculate $\delta_{it}$. In this equation, $q_{it}$ represents the cumulative percentage of residents fully vaccinated in county $i$ by week $t$. The numerator \eqref{eq:roc} represents the percentage of population vaccinated during the last $\ell$ weeks. The denominator  represents the percentage of unvaccinated in county $i$ in week $t-\ell.$ 

\begin{equation}\label{eq:roc}
\delta_{it} = \frac{q_{it} - q_{it-\ell}}{1 - q_{it-\ell}} \qquad \forall i \in C, \; t \in [4,43].
\end{equation}

Note that, $\delta_{it}$ measures the rate of change in vaccine uptake among the unvaccinated. Thus, $VH^b_{it} = 1 - \delta_{it} = \frac{1-q_{it}}{1-q_{it-\ell}}$ represents the fraction of unvaccinated that remained unimmunized during the period $t-\ell$ to $t$.  We use $VH^b_{it}$ as a comparative measure of VH behavior. This metric allows us to compare VH behavior among different counties at a particular point in time. For example, consider two counties, $i$ and $j$ that by period $t-\ell$ have vaccinated 60\% and 70\% of their population, correspondingly. Let us assume that during the last $\ell$ periods, both vaccinated 10\% of their population. As a result $\delta_{it} =\frac{10\%}{1-60\%} = 0.25$,  $\delta_{jt} =\frac{10\%}{1-70\%} = 0.33,$ and, $VH^b_{it} = 0.75$, $VH^b_{jt} =0.66$. The 10\% increase in vaccination leads to a higher vaccination rate for county $j$, and consequently a lower value of $VH^b_{jt}$. Let us consider a different example. Assume that two counties, $i$ and $j$ have vaccinated 60\% of their population by period $t-\ell.$ During the next $\ell$ periods, county $i$ vaccinates 10\% and county $j$ vaccinates 20\% of the population. In this case, $VH^b_{it} = 0.75$, $VH^b_{jt} =0.5$. County $j$ has higher vaccination rate, and lower value of $VH^b_{jt}.$

We calculated $VH^b_{it}$ for weeks 4 to 43, which correspond to  January 25, 2021 to October 31, 2021. We did not calculate $VH^b_{it}$ for the first 4 weeks of January 2021, although we have the data. This is because  vaccination delays during this period were mainly due to supply chain limitations rather than VH. After January 25$^{th}$ vaccines became available to everyone who wanted to get vaccinated, thus, vaccination delays were due to VH. The proposed metric may not be very effective when all counties are nearly fully vaccinated.  

Finally, let us highlight the differences among our proposed $VH^b$ and the estimates of VH attitude provided by the ASPE. ($i$) Our proposed $VH^b$ measure VH behavior, while ASPE's metric measures VH attitude. ($ii$) ASPE uses state-level data to derive the VH estimate at the county-level. More specifically, state-level VH estimates derived from surveys are converted  to Public Use Microdata Areas (PUMA) level estimates. Next, the PUMA-to-County crosswalk is used to generate county-level estimates of VH for week 31. These conversions may impact the accuracy of the estimates at the county-level. Different from ASPE, our metric uses county-level data. ($iii$) ASPE uses VH focused-surveys. Our proposed $VH^b$ uses several different sources of data, ranging from data related to economic, social, political factors, to social media, Google searches, etc. More importantly, ASPE'VH captures unwillingness to vaccinate, whereas our VH captures the change in unvaccinated between $t-\ell$ and $t$. 

{\bf Topic Modeling and Sentiment Analysis:} We used VADER to  assess the sentiment of a tweet. VADER computes a compound score that ranges from -1 to 1. A score of -1 represents an extremely negative sentiment, and + 1 represents an extremely positive sentiment. The compound score is calculated by summing the valence scores of each token in the lexicon. This score is adjusted according to multiple rules and then normalized. The valence score is a metric assigned to a word that gauges the sentiment of that word. For example, the valance score of \emph{``good''} is 1.9, and valance score of \emph{``horrible''} is -2.5. Additionally, there are three main follow-on actions available to the users in Twitter, namely retweet, like, and reply, whose counts indicate the tweet's exposure to the general public. Hence, the adjusted sentiment score factors in the number of likes and retweets associated with each tweet.

Although all retrieved tweets are related to COVID-19 in general, some of them can represent different topics/themes. Therefore, we investigate further to ensure their relevance in the study. For example, tweets regarding the Delta airline, were separated from tweets discussing the Delta variant of COVID-19. We used the Latent Dirichlet Allocation algorithm (LDA) to characterize topics of interest. LDA, from the \emph{Gensim} package available in Python, allowed us to discover hidden topics in an unsupervised manner \citep{blei_2003}. Our LDA model treated tweets as probabilistic distribution sets of words and topics. In other words, each tweet was viewed as a mix of multiple topics. Despite the usefulness of LDA, the outcomes can be challenging to interpret and can vary depending on the choice of hyperparameters such as $\alpha, \beta$, $K$. $\alpha$ represents document-topic density while $\beta$ represents topic-word density. $K$ is the desired number of topics to be reported by the LDA model. To select the best set of hyper-parameters, we conducted a grid search of all parameter combinations (i.e., $K$ varies from 5 to 30 with step size of 2, $\alpha$ and $\beta$ varies in \{0.01, 0.3, 0.6, 0.9, ``symmetric''\} with additional \{``asymmetric'', ``auto''\} for $\alpha$). The coherence score for each model was used to evaluate the quality of the topics. This score measures the semantic similarity of the words within a topic \citep{syed_2017}. Generally, the higher the coherence score, the better the topics extracted from the model. We selected the set of hyperparameters with the highest coherence score. The topic number with the highest percentage contribution in each tweet was assigned as the dominant topic. 

\begin{table}
    \TABLE
    {Summary of static and dynamic factors used in the RF model.\label{tab:factors}}
    {\begin{tabular}{@{}l@{\qquad}l@{\qquad}l@{}}
    \hline
    \multicolumn{2}{c}{\up Static Factors} & Dynamic Factors\down\\
    \hline\up
    Poverty Rate & Racial Composition & Weather PCs$^{(a)}$ \\
    Divorce Rate & Access to Internet & Stringency Index$^{(b)}$ \\
    Metro Status & Education Composition & Unemployment Rate$^{(c)}$ \\
    Diabetes Rate & Political Affiliation & Google Search Insights \\
    Life Expectancy & Vaccine Coverage Index & Tweets Sentiment Scores \\
    Age Composition & Percentage of Uninsured & Google Symptom Search Trends PCs$^{(a)}$ \\
    Labor Force Rate & Social Vulnerability Index & Average Online News Articles Tone \\
    Vaccination Sites & Historical Flu Vaccination &  \\
    Income per Capita & Healthcare Staff per Capita &  \down\\
    \hline
    \end{tabular}}
    {
    $^{(a)}$ Principle components. The number of PCs for weather, and Google symptom search trends are three and five, respectively\\
    $^{(b)}$ State level data\\
    $^{(c)}$ Monthly data
    }
\end{table}

{\bf Clustering:} 
 
\begin{figure}[h]
    \begin{center}
    \caption{Clustering counties using $VH^b$ during week 23 for every county in the CONUS: a) goodness of fit versus the number of clusters, b) distribution of $VH^b$ with natural breaks, and c) classified choropleth map of the clusters.}\label{fig:goodK}
    \includegraphics[width=.9\textwidth]{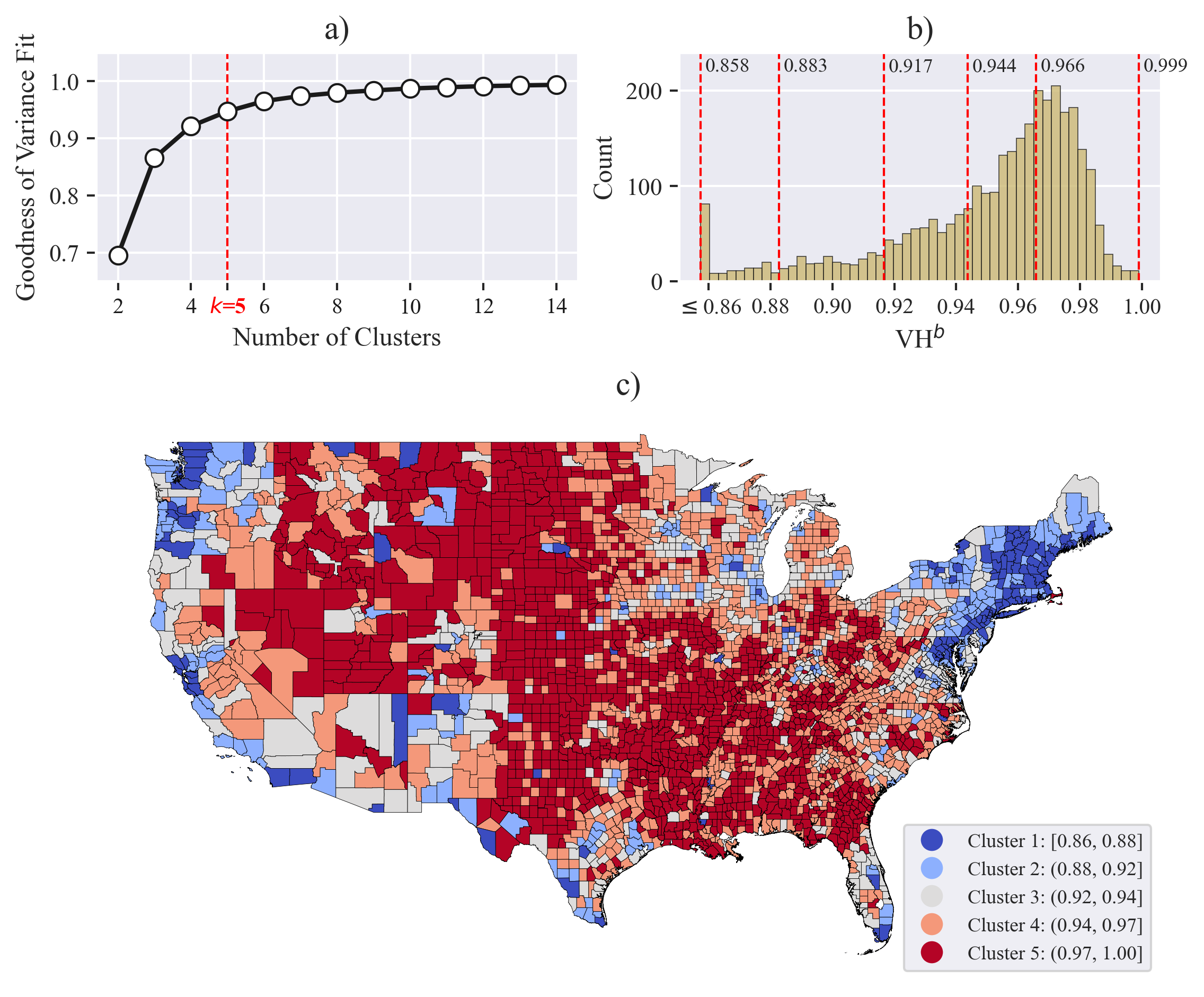}
    \end{center}
\end{figure}

We use the Fisher-Jenks algorithm, also known as the goodness of variance fit (GVF), to cluster counties based on $VH^b$ \citep{fisher_1958, jenks_1971, jenks_1977}. This algorithm minimizes the squared deviations of the class means. Figure \ref{fig:goodK} presents the outcome of this algorithm for week 23. We conducted a sensitivity analysis to evaluate the impact of the number of clusters on the GVF. Figure \ref{fig:goodK}a summarizes the results of this analysis. We use $k=5$ clusters, for which GVF is 95\%. Increasing the number of clusters beyond 5 does not provide a drastic improvement of GVF. The histogram in Figure \ref{fig:goodK}b presents  the distribution of $VH^b$ at the county level. 
For each cluster, we present a lower and upper bound of $VH^b$ (as shown via the red lines). Notice that,  traditional clustering methods, which use equal interval and quantile, would not provide a good classification due to the skewness of the $VH^b$ values. The choropleth map in Figure \ref{fig:goodK}c shows that the distribution of $VH^b$ across the U.S. is not even. About 71\% of the counties belong to clusters 4 and 5.

Large parts of the U.S., particularly in the north, central and southern regions, have high $VH^b$, which essentially means, it was hard for these counties to reduce the unvaccinated levels between $t -\ell$ and $t$. However, counties in coastal regions have lower overall $VH^b$. Counties which contain highly populated cities belong to \emph{C1} or \emph{C2}. Table \ref{tab:clusterStats} presents a few statistics for each cluster. \emph{C1} has the lowest $VH^b$ and only 3\% of the residents of \emph{C1} living in \emph{non-metro} counties. It also has the highest overall cumulative percentage of fully vaccinated residents (i.e., $54\%$). Thus, counties in \emph{C1} have progressed well in the race to vaccinate against COVID-19. On the other hand, \emph{C5} has the highest $VH^b$, and roughly half of its residents live in \emph{non-metro} counties.

\begin{table}
\begin{center}
    \TABLE
    {Cluster profile of the map shown in Figure \ref{fig:goodK}c (data of week 23).\label{tab:clusterStats}}
    {\begin{tabular}{@{}l@{\quad}l@{\quad}l@{\quad}l@{\quad}l@{}}
    \hline
    \up Cluster & \parbox[t]{2cm}{Average $VH^b$} & \parbox[t]{2cm}{Number of\\Counties} & \parbox[t]{2.1cm}{\% Population\\(Non-metro)} & \parbox[t]{2.7cm}{Average\\Vaccination Rate\down}\\
    \hline\up
    C1 & 0.830 & 142 & 0.033 & 0.540 \\
    C2 & 0.882 & 234 & 0.034 & 0.466 \\
    C3 & 0.921 & 511 & 0.080 & 0.406 \\
    C4 & 0.950 & 935 & 0.236 & 0.339 \\
    C5 & 0.973 & 1,243 & 0.577 & 0.271\down\\
    \hline
    \end{tabular}}
    {}
\end{center}
\end{table}

{\bf The RF Classification Model:} RF is one of the machine learning algorithms that has been widely used \citep{cutler_2012, fawagreh_2014, speiser_2019}. RF is an ensemble learning method used for classification and regression \citep{breiman_2001}. It constructs several decision trees using  bootstrap samples of training data, and random feature selection in tree induction. The RF classification model selects the best solution based on the majority vote (across the trees in the ensemble) for the class label. 

We use permutation and SHAP values to determine feature importance. We use the \emph{scikit-learn} package in Python to calculate the permutation importance values. This approach shuffles feature's values,  and the corresponding reduction in the model's performance is measured. The feature is important if after shuffling, the model's error increases. This approach provides a global insight of the model's behavior. In contrast, the Shapley Additive Explanations (SHAP) method  provides local explanations of the prediction made by the RF classification model \citep{shap_2020}. This is a model-agnostic method, since it can be used by any ML model.  The SHAP method computes the Shapley values which is a concept from  cooperative game theory. It quantifies the contribution of each feature  to the output of the RF model. In our application, the output of the RF model is the predicted probability that a county belongs to a particular cluster. We calculate these  probabilities by dividing the number of votes for each cluster by the number of trees in the forest. 

Based on our experiments, we notice that the ranking of feature importance depends on the approach used (i.e., permutation of feature importance and SHAP value). However, the top 15 features identified by both methods are similar.

\subsection{Model Validation and Verification}
{\bf $VH^b$:} To evaluate whether $VH^b$ is a relevant measure of VH behavior, we compare its values with VH estimates calculated using data from two large-scale surveys, one conducted by the ASPE, and the other conducted via the collaboration between Delphi group at Carnegie Mellon University and Facebook (referred to as Delphi in this paper).

\begin{figure}[ht!]
    \FIGURE
    {\includegraphics[width=0.9\textwidth]{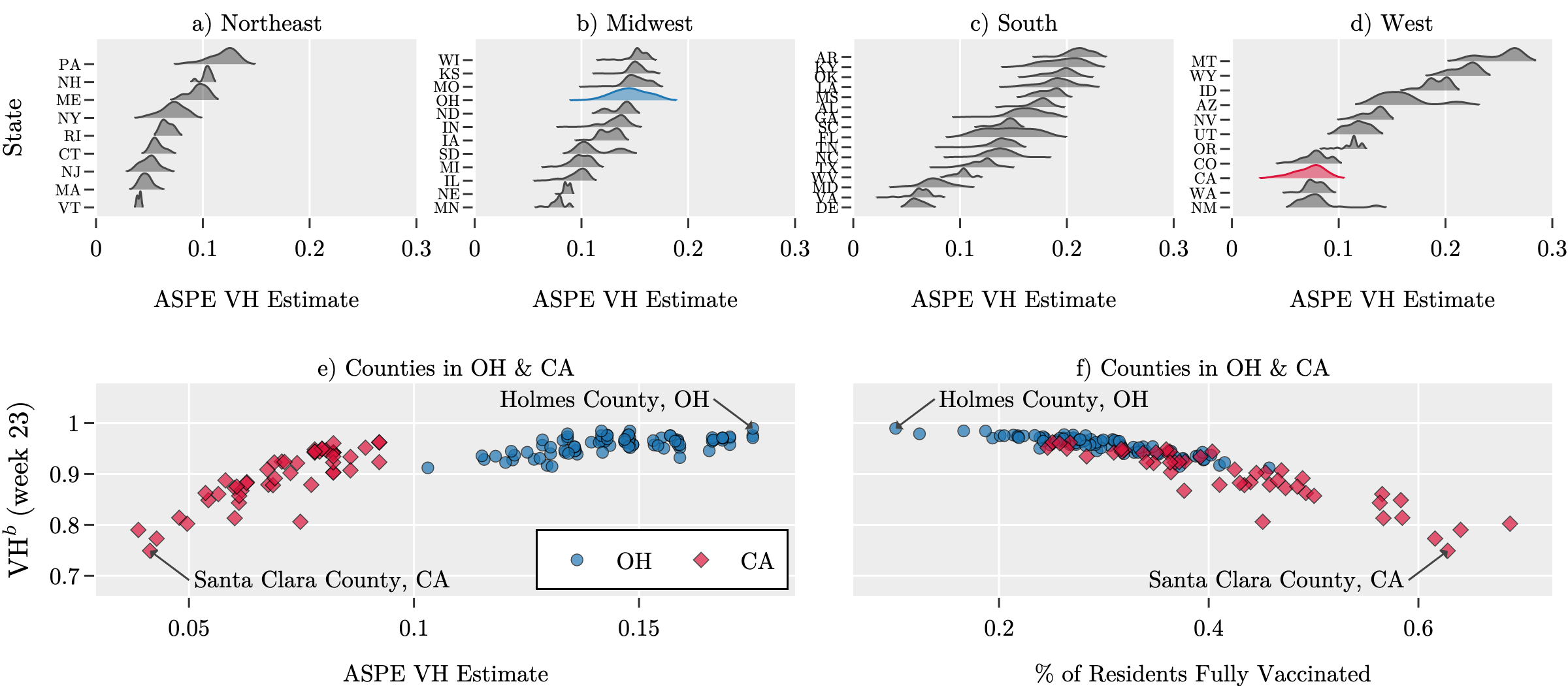}}
    {(a)-(d) Distributions of the VH estimates among counties in each State of USA using data from the ASPE \citepalias{aspe_vh}. (e) Relationship between the $VH^b$ and ASPE's VH estimate during week 23; and (f) relationship between the percentage of fully vaccinated residents and $VH^b$ for counties in Ohio and California. Each marker represents a county.\label{fig:hesSta}}
    {}
\end{figure}

The ASPE's surveying was conducted during May 26 to June 7, 2021 (weeks 21-23). Figures \ref{fig:hesSta}(a) to (d) summarize the estimates of VH index at the county level in different states of the U.S. reported by ASPE \citepalias{aspe_vh}. We notice that  VH index is low in the Northeast region. The VH index in a few states in the South and West are greater than 0.2.  Figure \ref{fig:hesSta}(e)  compares ASPE's VH index and $VH^b$ for counties in California and Ohio. This graph shows a positive relationship between these measures.  Figure \ref{fig:hesSta}(f) compares the percentage of residents fully vaccinated and $VH^b$. This graph shows a negative relationship between these measures, indicating that counties with higher level of vaccination have lower $VH^b$.  

\begin{figure}[ht!]
    \FIGURE
    {\includegraphics[width=1\textwidth]{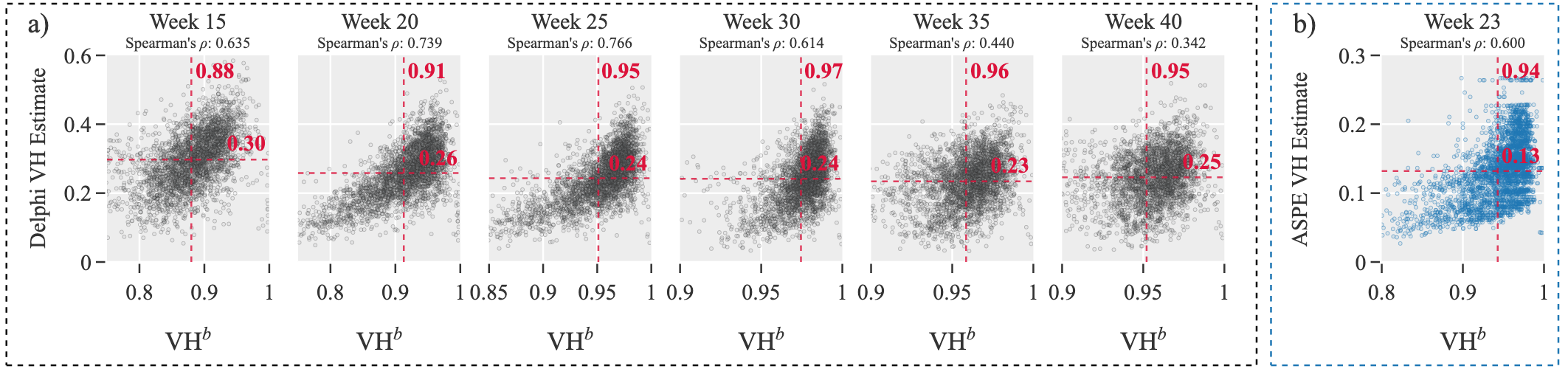}}
    {(a) Relationship between $VH^b$ and VH estimate by Delphi group and Facebook, for various weeks. (b) Relationship between ASPE's VH and $VH^b$ estimates in week 23.\label{fig:hesDel}}
    {}
\end{figure}

Figure \ref{fig:hesDel}(a) presents the relationship between $VH^b$ and  VH estimate from  Delphi for several weeks during the period of study. Notice that, estimates of VH via Delphi  are available every week, at the county level, beginning January 2021. We present the data for weeks 15, 20, \ldots, 40. Figure \ref{fig:hesDel}(b) provides the relationship between $VH^b$ and ASPE's VH estimate for week 23. Each dot in these figures represents a county. The red dash line represents the mean of the observations. The values of the correlation coefficients between $VH^b$ and Delphi's VH estimates vary between 0.34 and 0.75. The value of the correlation coefficients between $VH^b$ and ASPE's VH estimate is 0.58. These results indicate a positive relationship. That means, in general, counties that have low $VH^b$, do also have low values of ASPE's VH and Delphi's VH estimates, and vice versa. This indicates that our proposed $VH^b$ is an effective tool to predict VH behavior at the county level. 

Notice that, the average Delphi estimate of VH decreases from 0.30 to 0.25 from week 15 to 40. This corresponds to a 17\% reduction of VH during 23 weeks. This change in attitude toward vaccination  could be the outcome of vaccination mandates employed at the state level, community outreach, etc. 

\begin{figure}[ht!]
    \FIGURE
    {\includegraphics[width=0.9\textwidth]{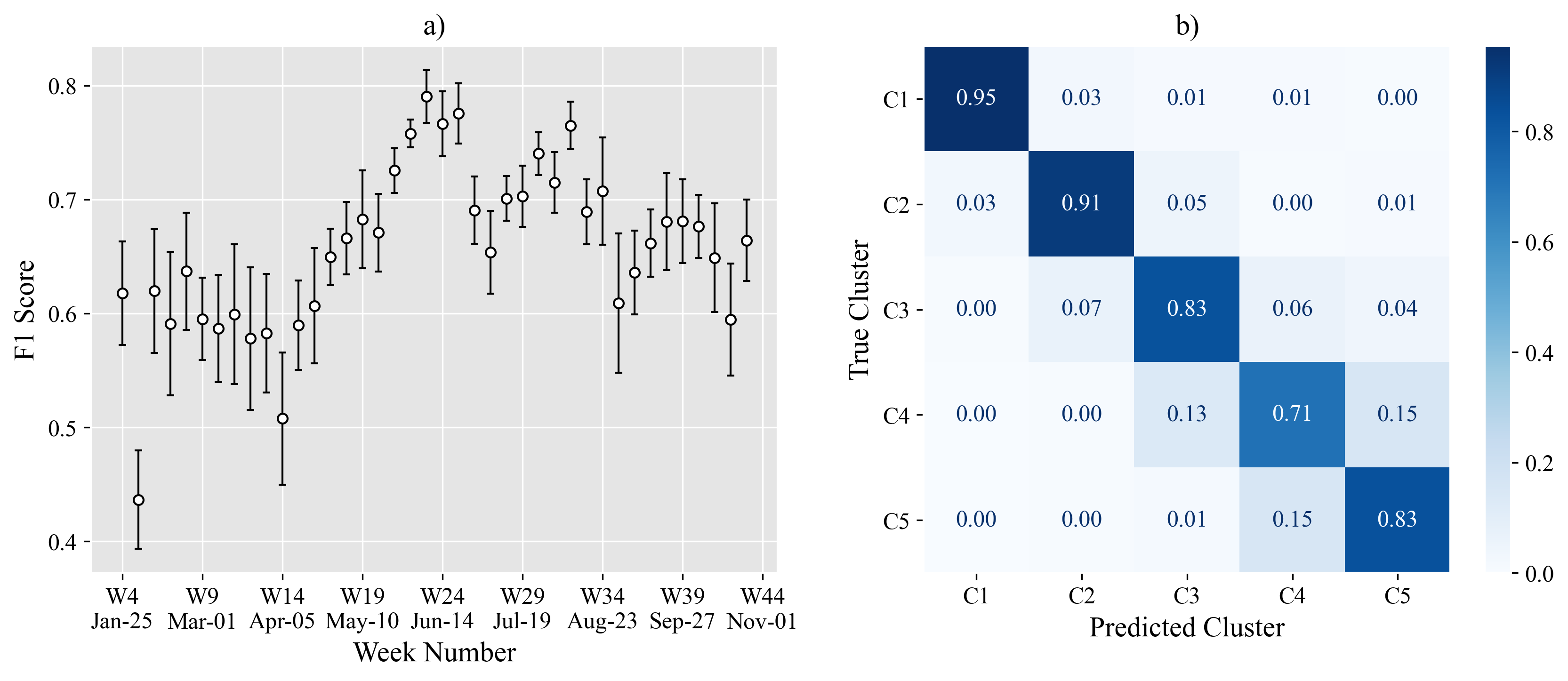}}
    {(a) Average $\pm$ 1 standard deviation of macro F1-score; and (b) Normalized confusion matrix of cluster predictions using data of week 23.\label{fig:rfPerf}}
    {}
\end{figure}

{\bf RF Classification Model:} We use the data collected to develop an RF classifier model for each week during the period of study to determine the most relevant factors to predict the cluster labels. We use a 5 fold cross-validation (5-CV) to train and validate the model. Next, we calculate the F1-score, the harmonic mean of precision and recall. We use F1-score, rather than precision or recall, as a performance measure of RF classifier, since we assume that errors caused by false positive, or false negative classifications to have the same importance. Figure \ref{fig:rfPerf} summarizes the macro F1-scores for weeks 4 to 43. For each week, we have present the average $\pm$ 1 standard deviation of the macro F1-score calculated from model training via 5-CV. The results show that the lowest F1-scores occur during weeks 4-14. This is mainly due to errors, incomplete and inconsistent data during the early stages of data collection, which were because of changes in the content of the data reported to CDC in late February 2021 \citepalias{cdc_data_updates}, and the differences in vaccine roll out plans adopted at the state level.

Figure \ref{fig:rfPerf}b presents the normalized confusion matrix for the RF classifier of week 23, which has one of the highest F1-scores. The values of the diagonal elements represent the degree of correctly classified values. The diagonal values for \emph{C1} and \emph{C2} are above 90\%, which indicate that the model classifies with high  accuracy whether a county belongs to these two clusters. Based on these values, model's  performance is moderately accurate for \emph{C3} and \emph{C5}. Model's performance is worst for \emph{C4}. The relatively worst performance of the model in classifying counties that belong to \emph{C4} and \emph{C5} is because of the large size of these clusters as compared to the rest. The size of these clusters leads to imbalances.

\begin{figure}[htp!]
    \FIGURE
    {\includegraphics[width=0.7\textwidth]{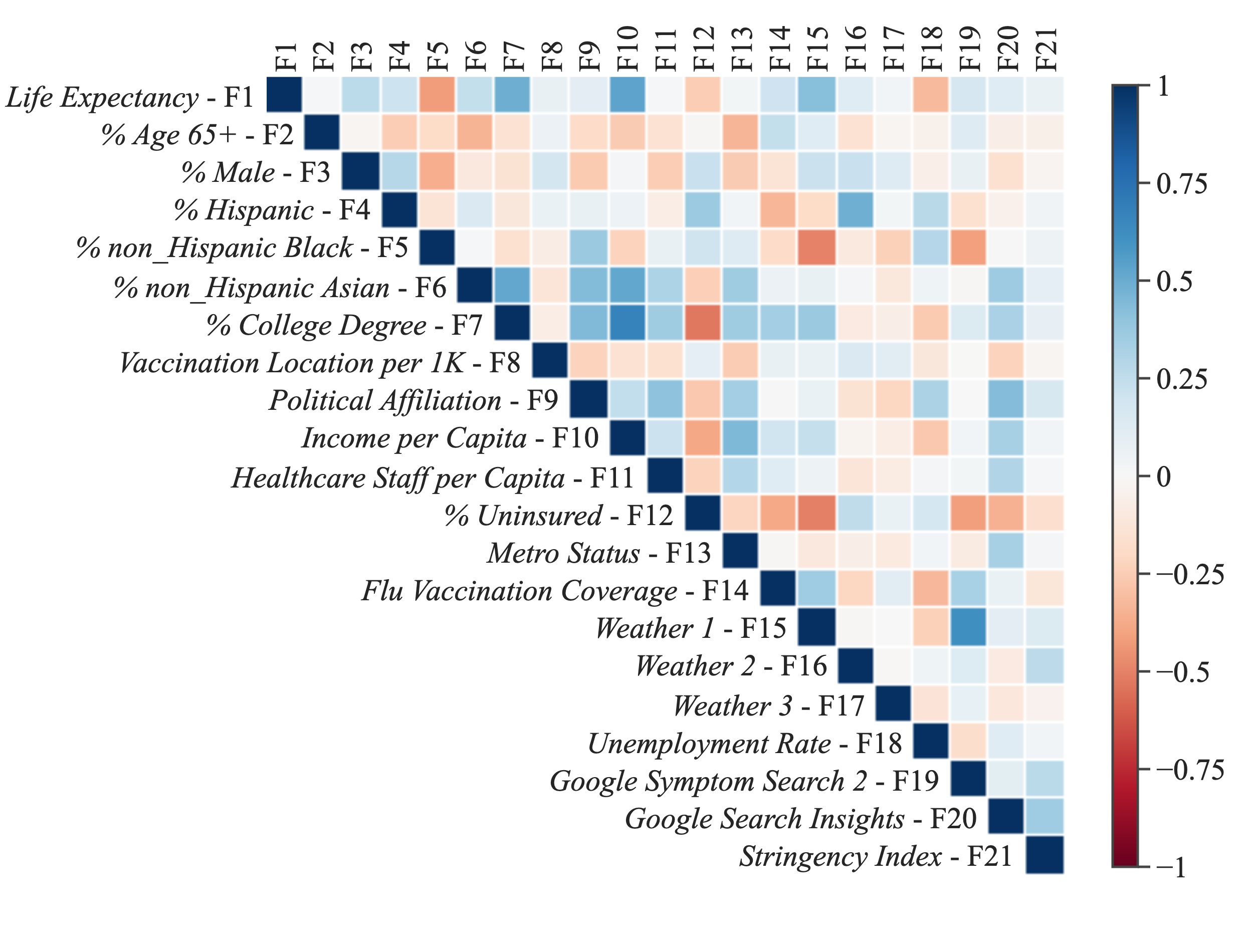}}
    {Heatmap of the correlation matrix across the features after filtering out the highly correlated features. For ease of visualization, Tweet sentiment features for different topics are not shown as they do not exhibit a high correlation with other features.\label{fig:corrHeatmap}}
    {}
\end{figure}

{\bf Multicollinearity Effects:} 
Figure \ref{fig:corrHeatmap} depicts the Pearson's correlation heatmap for the features we include on the RF regression. The correlation coefficients ($r$) for \emph{\% of College Degree} and \emph{Income per Capita} is $r = 0.676$; for \emph{Weather 1} and \emph{Google Symptom Search 2} $r = 0.620$; and for \emph{\% of College Degree} and \emph{\% Uninsured} $r = -0.533$. The correlation coefficient for the rest of the features is smaller than $\lvert 0.5 \rvert$. 

A similar analysis of other features in our aggregated dataset showed that \emph{\% High School} and \emph{\% College Degree} are strongly negatively correlated, and $r=-0.783$; \emph{Income per Capita} and \emph{\% Below Poverty} are strongly negatively correlated, and  $r=-0.747$. We dropped a few features from the RF regression model which are highly correlated with other features in order to reduce the size of the RF classifier.  

\section{Discussion of Results}
\label{sec:discussion}
Via our discussion of results we address the research questions identified in Section \ref{sec:intro}.

\begin{figure}[t]
    \FIGURE
    {\includegraphics[width=0.9\textwidth]{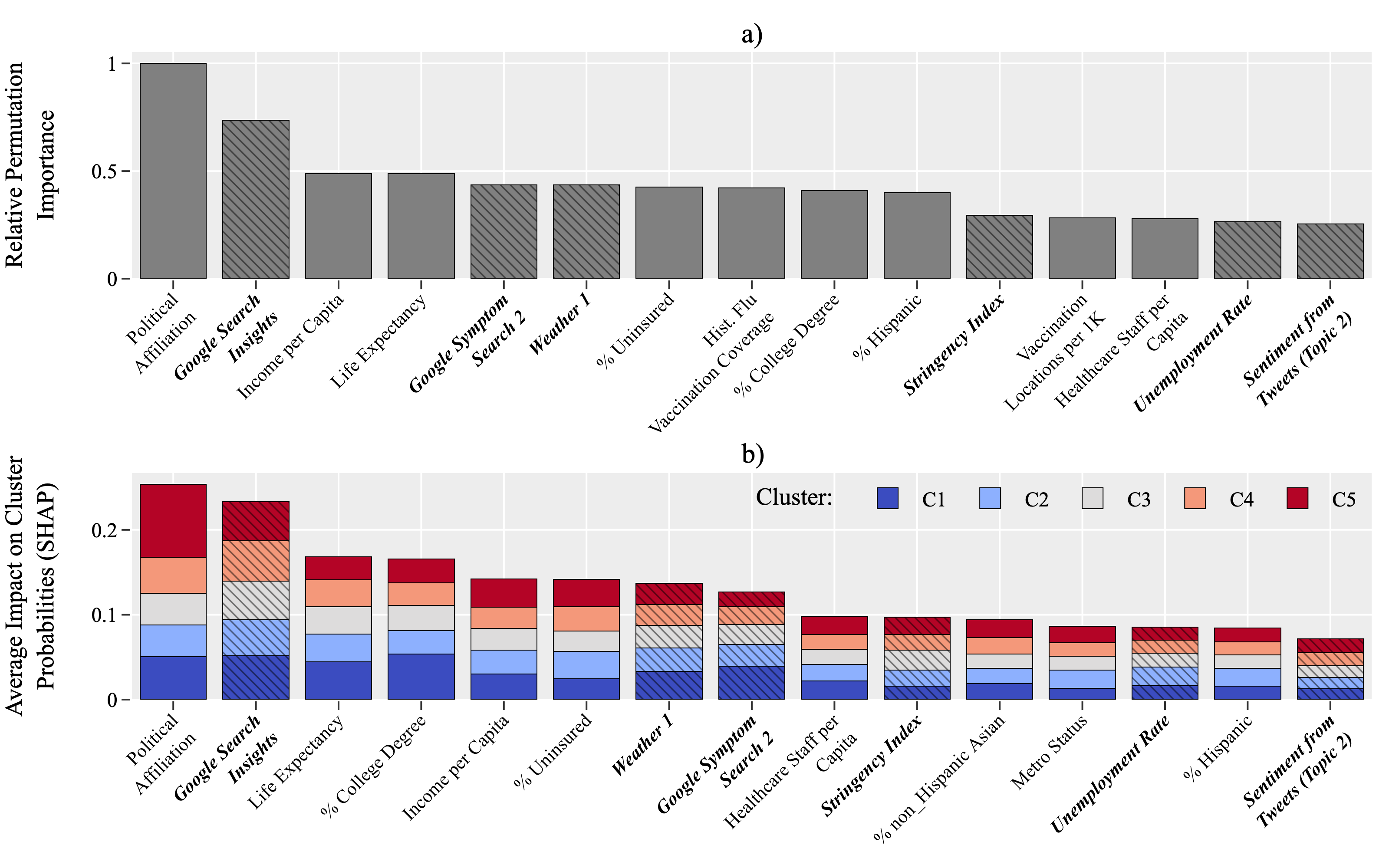}}
    {(a) Permutation feature importance of the top 15 relevant features, (b) SHAP feature importance of the top 15 relevant features. In both graphs, bars with hashed patterns and bold labels represent dynamic factors. The results derive from the classifier using the inputs of week 23.\label{fig:importance}}
    {}
\end{figure}

{\bf What factors impacted $VH^b$ for COVID-19 vaccine?} Figure \ref{fig:importance}a presents the permutation importance value of the top 15 relevant features of the model during week 23. Figure \ref{fig:importance}b presents the SHAP values of the top 15 features that have the most impact on the model output. In both plots, the bars with hashed patterns represent the dynamic factors.  We discuss the findings for week 23 because the predictions of the RF classification model for this week are highly accurate (the F-1 score is the highest, see Figure \ref{fig:rfPerf}).

Both approaches indicate that Political Affiliation is the most influential factor in predicting VH behavior of the population in a particular county. Here, Political Affiliation presents the percentage of people in a county that voted for the Democratic candidate during the 2020 presidential elections. This result aligns with a similar finding discussed in a New York Times article \citep{nyt_political}. This article finds a strong correlation between the distribution of votes among political parties during the 2020 presidential elections and VH. The data used in this study to estimate VH was collected from a survey. 

The feature related to the trend of Google searches for COVID-19 vaccination information is found to be the next important feature. Here, Google Searches refers to the aggregated (and anonymized) trends in Google searches  related to COVID-19 vaccination \cite{google_vsi_description}. The healthcare staff per capita, unemployment rate, and metro status were among the least relevant features in determining the $VH^b$ of a county. 

Based on our analysis, age is not an important factor to determine VH, although CDC indicates that unvaccinated older adults are more likely to be hospitalized and die from COVID-19 \citepalias{CDC_olderAdults}.  However, since vaccination of older adults was prioritized, many were vaccinated as soon as COVID-19 vaccines were made publicly available (December 2020 to January 2021). Our dataset does not include this time period, which may explain this observation.        

\begin{figure}[t]
    \FIGURE
    {\includegraphics[width=0.7\textwidth]{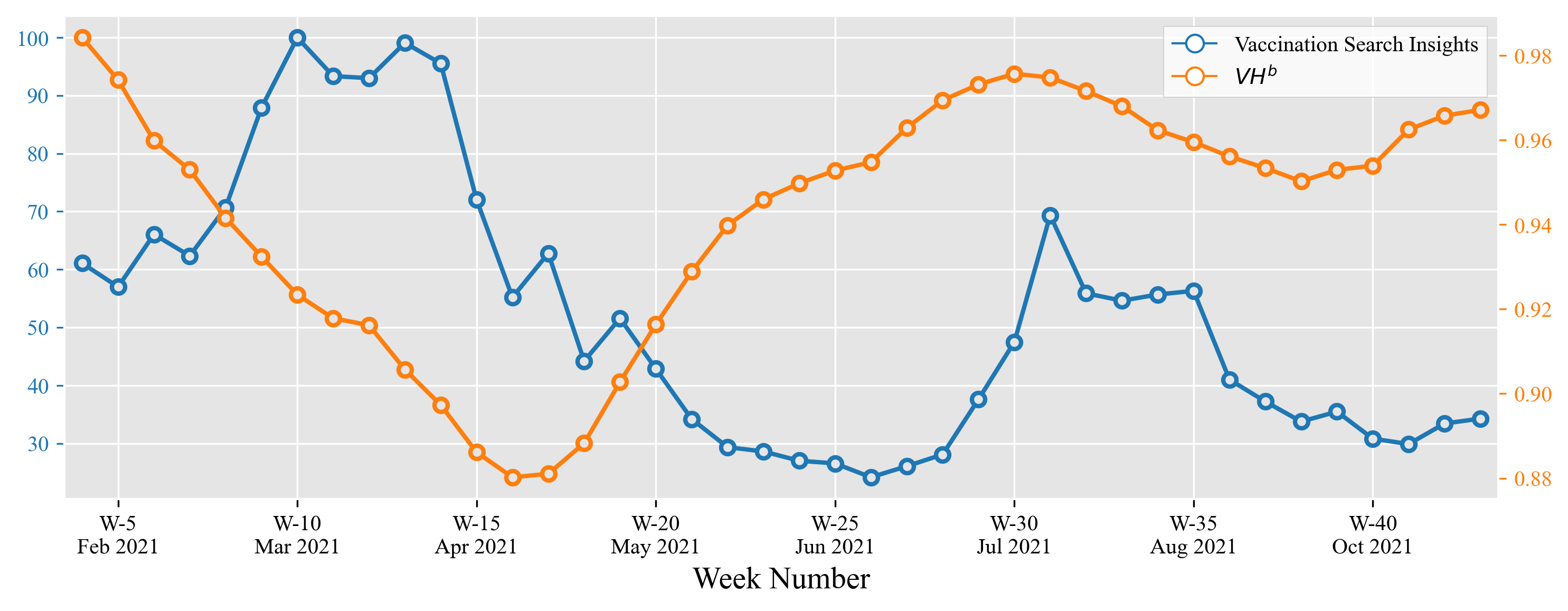}}
    {Overall $VH^b$ and Google search trends over time in the U.S.\label{fig:search_insights}}
    {}
\end{figure}

Although the static factors (such as political affiliation, education level, income per capita, etc.) make up the majority of relevant factors; we notice that a few dynamic factors are relevant. Since the values of dynamic factors change over time, they can help explain changes we observe in VH behavior. For example, Figure \ref{fig:search_insights} presents the values of $VH^b$ and Google's Vaccination Search Insight \citep{googleins}. Vaccination Search Insight represents the number of relative (to other participating countries) Google searches related to eligibility and accessibility of COVID-19 vaccines. The graph demonstrates that Google search trends can explain some of the changes we observe in VH behavior  over time.    

Our model recognizes the Stringency Index as a relevant factor. The Stringency Index records the strictness of ``lockdown style'' policies, which restrict people's behavior \citep{SI_trucker}. The index is a composite score from nine indicators ranging from school closures to public information campaigns. Although the model determines the Stringency Index to be a relevant factor, it is worth mentioning that it is challenging to determine how and when it impacted vaccination uptake. This is mainly because there is a lag between the time a policy is implemented and the time its impacts are observed. Additionally, several policies are implemented at the State or Federal level; however, we observe vaccine uptake and $VH^b$ at the county level. Hence, the outcome of these policies at the county level is impacted by other factors.

Another relevant dynamic factor is the sentiment of Topic 2. This topic contains tweets related to people's emotions about COVID-19. The top ten keywords included in this topic are ``year'', ``family'', ``miss'', ``pray'', ``friend'', ``thank'', ``love'', ``old'', ``lose'', ``wish''. These words explain people's perceived risk from COVID-19, and can help explain VH behavior. There are two possible explanations for why the sentiments of other topics are irrelevant. First, the limited number of tweets with geographical metadata might have restricted our opportunity to capture the sentiments about other topics. Secondly, noise in the data and viral tweets/memes do not contribute any value to the topic. 

\begin{figure}[t]
    \FIGURE
    {\includegraphics[width=1\textwidth]{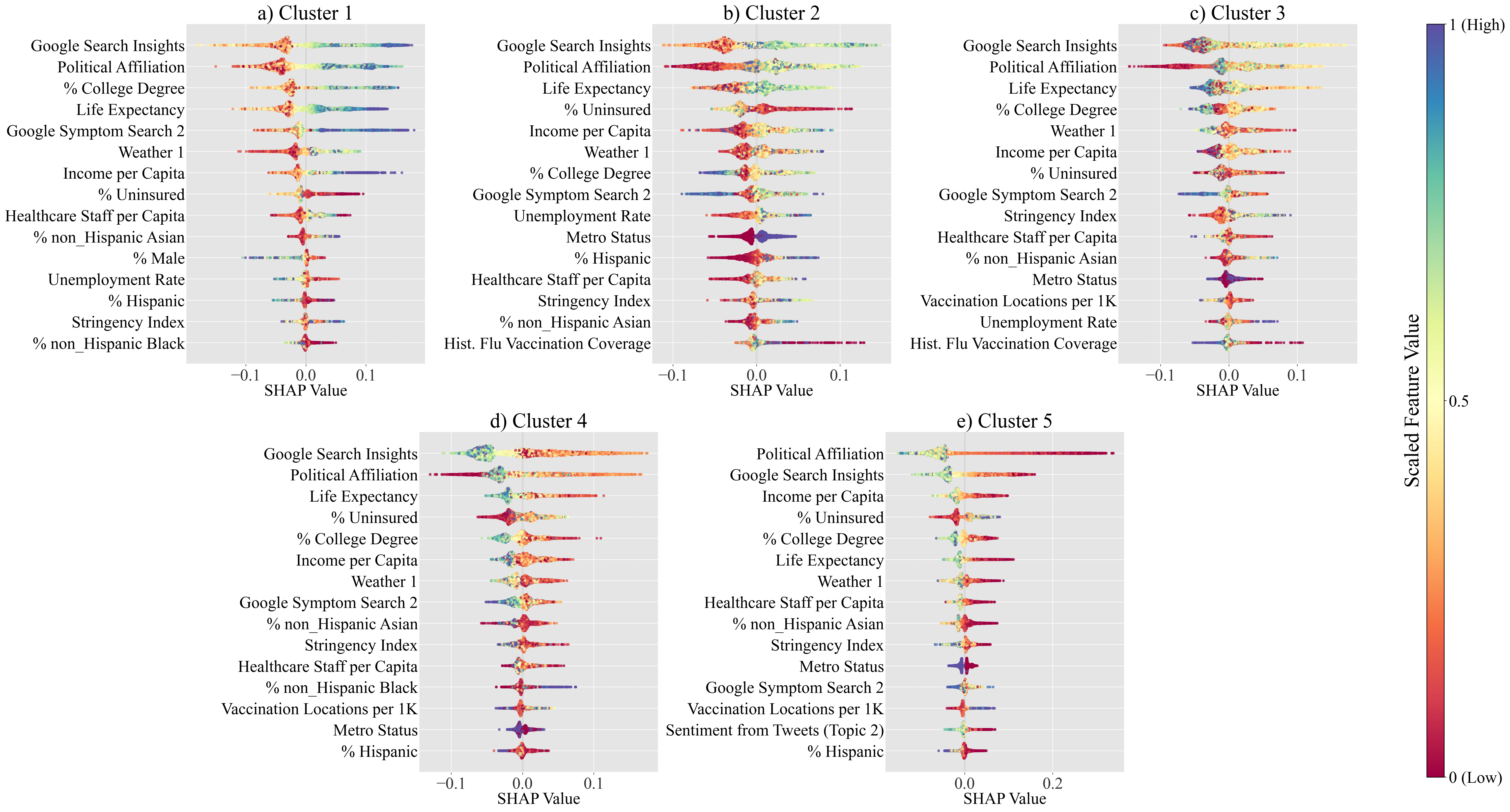}}
    {SHAP values of the top 15 important features for each cluster.\label{fig:shap}}
    {}
\end{figure}

{\bf In what  meaningful clusters should counties be aggregated to support efforts of overcoming VH?}
The trained RF classification model aggregated counties of the U.S. into 5 clusters, \emph{C1},\ldots, \emph{C5}.
Figure \ref{fig:shap} presents the SHAP values for the most important features of these clusters in week 23.

In Figure \ref{fig:shap}, for each cluster, features are sorted in decreasing order of the total (over the counties) SHAP value magnitudes. Table \ref{tab:clusterProfi} presents a summary of the data displayed in Figure \ref{fig:shap}. The table presents the average SHAP value  of the 15 top features of each cluster.  Based on these results, Google Search Insight  and Political Affiliation are the most relevant features across all clusters. 

\begin{table}
\TABLE
{The average feature value per predicted cluster of the top 15 important features.\label{tab:clusterProfi}}
{\begin{tabular}{@{}r@{\quad}l@{\quad}l@{\quad}l@{\quad}l@{\quad}l@{}}
\hline
\up Predicted Cluster Label & C1 & C2 & C3 & C4 & C5\down \\
\hline\up
Number of Counties & 141 & 240 & 519 & 928 & 1,237 \\
Total Population & 56.5M & 79.4M & 96.7M & 61.9M & 30.8M \\
Average $VH^b$ & 0.829 & 0.884 & 0.922 & 0.951 & 0.972 \\
\hline\up
Top 15 Features & \multicolumn{5}{c}{Average Feature Value\down} \\
\hline\up
Political Affiliation & 0.60 & 0.48 & 0.42 & 0.33 & 0.24 \\
Google Search Insights $^a$ & 33.71 & 27.83 & 23.94 & 19.09 & 15.45 \\
Life Expectancy & 80.45 & 79.25 & 78.28 & 77.41 & 77.06 \\
\% College Degree & 0.45 & 0.38 & 0.34 & 0.30 & 0.27 \\
Income per Capita & \$ 78,184 & \$ 64,988 & \$ 57,401 & \$ 51,749 & \$ 47,630 \\
\% Uninsured & 0.07 & 0.08 & 0.10 & 0.12 & 0.14 \\
Weather 1 $^a$ & 1.28 & 0.81 & 0.23 & -0.13 & -0.37 \\
Google Symptom Search 2 $^a$ & -0.58 & -0.83 & -0.92 & -0.96 & -0.96 \\
Healthcare Staff per Capita & 0.69 & 0.58 & 0.49 & 0.40 & 0.31 \\
Stringency Index $^a$ & 41.23 & 40.60 & 39.86 & 37.66 & 36.71 \\
\% non\_Hispanic Asian & 0.06 & 0.03 & 0.02 & 0.01 & 0.01 \\
Metro Status (1=Metro) & 0.72 & 0.70 & 0.56 & 0.39 & 0.19 \\
Unemployment Rate $^a$ & 0.05 & 0.06 & 0.06 & 0.05 & 0.05 \\
\% Hispanic & 0.11 & 0.13 & 0.10 & 0.08 & 0.08 \\
Sentiment from Tweets (Topic 2) $^a$ & 0.33 & 0.32 & 0.32 & 0.31 & 0.24\down\\
\hline
\end{tabular}}
{$^a$ Dynamic feature}
\end{table}

Based on our model, people who live in counties of \emph{C1} showed the least resistance to getting vaccinated (the value of $VH^b$ is lowest at 0.829). Most of the people who live in these counties are democrat, are  more internet-inquisitive (are more prone to seek information from multiple sources in the internet), have the longest life expectancy, have a college degree, have the highest income per capita, have the lowest percentage of uninsured, live  in metropolitan areas, live in areas with the highest number of healthcare staff per capita, and the highest Stringency Index.

People who live in counties of \emph{C5} showed the highest resistance to vaccination. Most of the people who live in these counties are republicans, are the are the least internet-inquisitive, (are more prone to seek information from multiple sources in the internet), have the shortest life expectancy, do not have a college degree, have the lowest income per capita, have the highest percentage of uninsured, live  in non-metropolitan areas, live in areas with the lowest number of healthcare staff per capita and the lowest Stringency Index. We leave it to the reader to discern the remaining clusters and features.

\begin{figure}[t] 
    \FIGURE
    {\includegraphics[width=1\textwidth]{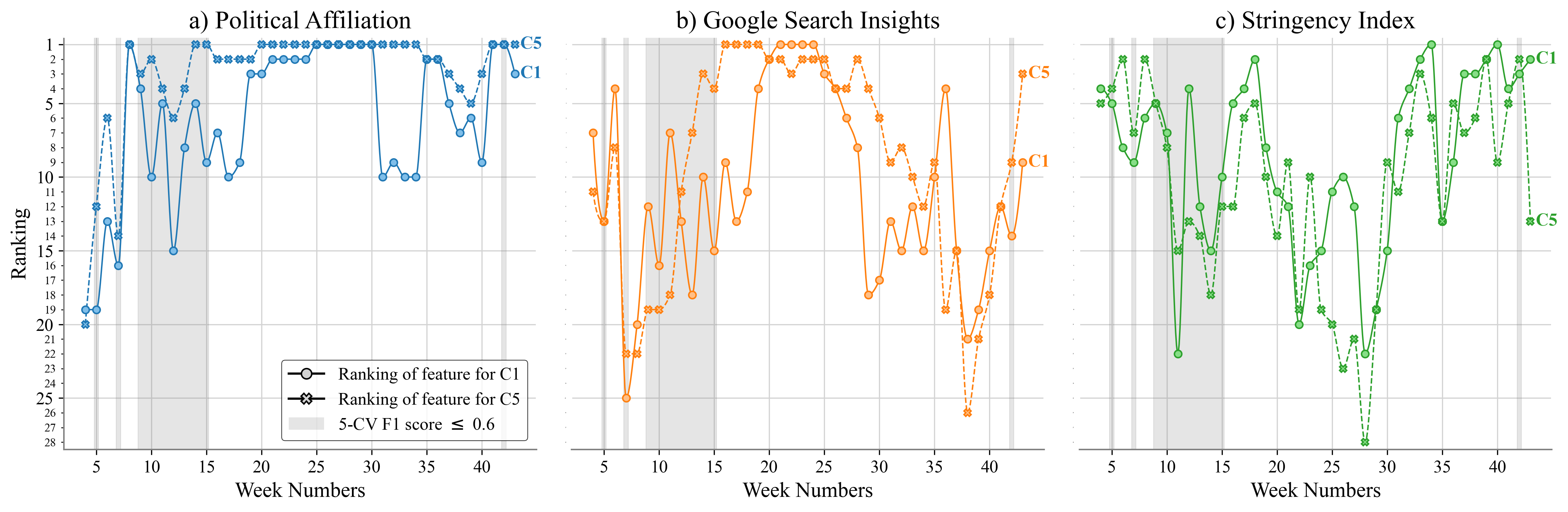}}
    {Ranking of political affiliation, Google search insights, and Stringency Index over time for \emph{C1} and \emph{C5}.\label{fig:stringency_political}}
    {}
\end{figure}

{\bf How did vaccination-related policies and COVID-19 restrictions impact VH in the U.S.?}
Figure \ref{fig:stringency_political}  presents the ranking of political affiliation, Google search insights, and Stringency Index over time for clusters \emph{C1} and \emph{C5}. The RF classification models for weeks in shaded areas have the lowest accuracy (F1-score is below 0.6). The  results of Figure \ref{fig:stringency_political}a  indicate that, for \emph{C5}, political affiliation was the most relevant factor during 21 weeks, and the second most relevant factor during 7 weeks. Political affiliation was more important to \emph{C5} (as compared to \emph{C1}) during 31 weeks, and was of the same importance during 9 weeks (out of 41 weeks of study). These results show that political affiliation for \emph{C5} dominates the other factors. Based on these results,  political affiliation was a very important factor in deterring the population of \emph{C5} from getting vaccinated.  

Based on the results of Figure \ref{fig:stringency_political}b, Google search insights was most relevant in determining $VH^b$ of \emph{C5} (as compared to \emph{C1}) during 22 weeks, and was of the same relevance during 5 weeks. Google search insights was most relevant in determining $VH^b$ of \emph{C1} (as compared to \emph{C5}) during 14 weeks. The SHAP values for week 23 (Figure \ref{fig:shap}) show that counties that have the highest number of Google searches most probably belongs to \emph{C1}, and counties with the lowest number of Google searches most probably belong to \emph{C5}. In summary, the population living in counties that belong to \emph{C5} had less interest to search on-line about the eligibility and accessibility of COVID-19 vaccines than those in \emph{C1}. This factor may explain low vaccination uptake (higher $VH^b$) of \emph{C5}.   

Based on the results of Figures \ref{fig:stringency_political}c   Stringency Index was the most relevant feature for \emph{C1} during 2 weeks. Stringency Index was never found to be the most relevant feature for \emph{C5}. The Stringency Index was more important to \emph{C1} (as compared to \emph{C5}) during 25 weeks, and was of the same importance during 4 weeks. Stringency Index has the greatest impact on $VH^b$ during week of May 3rd (week 18), August 16th (week 33), August 23rd (week 34), September 27th (week 39), and October 4th (week 40).  

\begin{figure}[t]
    \FIGURE
    {\includegraphics[width=0.8\textwidth]{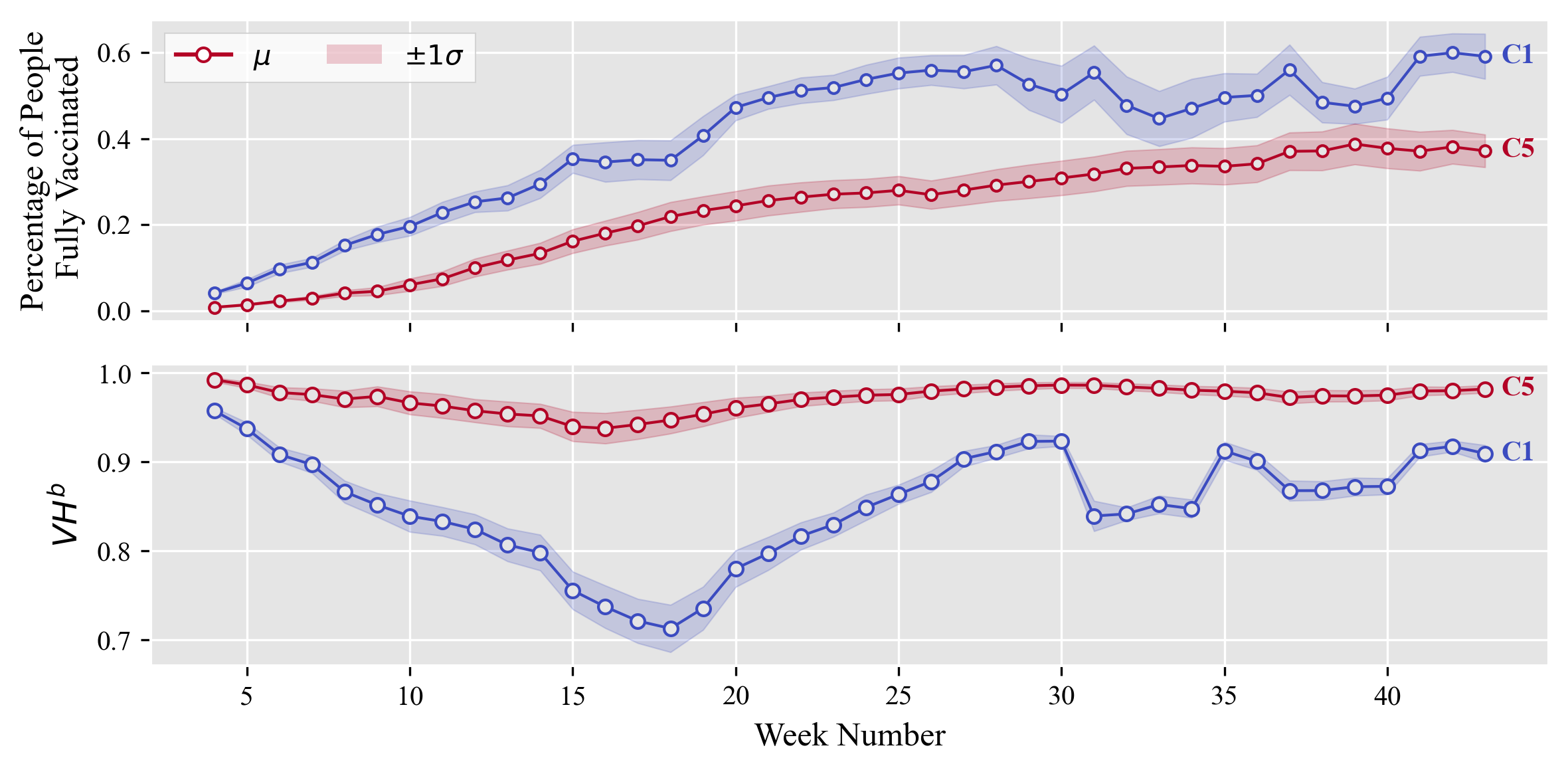}}
    {The average $\pm$ one standard deviation of percentage of fully vaccinated and $VH^b$ over time in \emph{C1} and \emph{C5}.\label{fig:stringency}}
    {}
\end{figure}

We further investigate the role of the Stringency Index on $VH^b$. Note that, the Stringency Index is comprised of nine indicators, some of which are school closing, workplace closing, canceling public events, etc. Recall that, on July 27, 2021, CDC announced an upswing in cases due to the Delta variant. As a result, several States recommended that people avoid travel  to reduce the spread of the virus. Due to the outbreak of the Delta variant, and the quick spread of the disease during the first weeks of the Fall semester, several school districts shut down in-person classes. The disease outbreak affected State policies and interventions, which in return  encouraged people to get vaccinated. This stream of events seems to have had a greater impact on increasing vaccination uptake in \emph{C1} rather than \emph{C5}. Notice the changes in the average ($\pm$ one standard deviation) percentage of fully vaccinated and $VH^b$ of \emph{C1} and \emph{C5} during weeks 31 to 41 in Figure \ref{fig:stringency}. The value of $VH^b$ for \emph{C5} does not change much during these weeks. As a result, the average percentage of fully vaccinated increases steadily, and at a lower rate than in \emph{C1}. However, the values of $VH^b$ for \emph{C1} change drastically. We also observe changes in the average percentage of fully vaccinated during weeks 31 to 41.

In Figure \ref{fig:stringency} we observe a decreasing trend of $VH^b$  and an increasing trend of vaccination uptake of \emph{C1} during weeks 4 to 18. During this period, the supply chain of COVID-19 vaccines faced several challenges \citep{bollyky_2021}. Thus,  people were vaccinated gradually as vaccines became available.  By the end of April 2021 (week 18), vaccines were available to everyone. Thus, the changes observed in $VH^b$ are partly due to VH. We observe an increase of $VH^b$ of \emph{C1} during weeks 19 to 30. This does not necessarily mean that people are becoming resistant to vaccination. Since most people are already vaccinated, the rate at which people are vaccinated is reduced. The ability to increase marginal gains in immunization is affected. In summary, $VH^b$ presents relative changes of VH behavior over time. One can use $VH^b$ to compare VH behavior of different populations over time to measure relative resistance to immunization.

\section{Summary of Results and Conclusions}\label{sec:conclusions}

{\bf Summary of the Proposed Research:} This research proposes a modeling framework that fuses rich static and dynamic datasets (via a  machine learning (ML) algorithm) to  explain why people are hesitant to get the COVID-19 vaccine in the U.S. We collected a vast amount of data from different sources during the period of January to October 2021.  We propose a simple metric of vaccine hesitancy (VH) behavior, $VH^b$, which characterizes hesitancy as marginal gain in immunization over time. We compare $VH^b$ to VH estimates provided via data collected from surveys. The ML algorithm is a Random Forest (RF) classification model that is simple and flexible for incorporating new features with little effort. We train and validate the model using a 5-fold cross-validation procedure. We use the SHAP values to measure the impact of features on the model output. The model groups the counties of the U.S. into 5 major clusters. For each cluster, we determine the most relevant factors and provide a discussion to explain their VH behavior.   

{\bf Research Findings:} We make the following observations:\par ($i$) We propose a comparative measure of the VH behavior, $VH^b$. It presents relative changes of VH behavior over time. One can use $VH^b$  to compare VH behavior of different groups over time. 
\par
($ii$) Google search insights and political affiliation were the most relevant features in determining VH at  the county level. \par ($iii$) Dynamic features, such as, Google searches related to COVID-19, Stringency Index, weather, unemployment rate, Tweet sentiments were found relevant to explain dynamic changes of VH over time at the county level.\par  ($vi$) Most of the population in counties with the least resistance to getting vaccinated (cluster \emph{C1}) are  democrat, are more internet-inquisitive (are more prone to seek information from multiple sources in the internet), have the longest life expectancy, have a college degree, have the highest income per capita, have the lowest percentage of uninsured, live in metropolitan areas, live in areas with the highest number of healthcare staff per capita and the highest Stringency Index. \par ($v$) Most of the population in counties with the highest resistance to getting vaccinated (cluster \emph{C5}) are republicans, are  the least internet-inquisitive, (are least prone to seek information from multiple sources in the internet), have the shortest life expectancy, do not have a college degree, have the lowest income per capita, have the highest percentage of uninsured, live  in non-metropolitan areas, live in areas with the lowest number of healthcare staff per capita and the lowest Stringency Index. \par ($vi$) Vaccination-related policies and COVID-19 restrictions, as measured by the Stringency Index, were effective in increasing vaccination uptake of counties in cluster \emph{C1}. These policies and restrictions did not seem to be effective in counties that belong to cluster  \emph{C5}. \par

{\bf Research Limitations:} Prediction accuracy of our proposed model and the corresponding outcomes are affected  by:\par 
($i$) \emph{Quality of data used}: The quality of the data collected differs across counties in the U.S. We are missing data about the vaccination uptake, Tweets, etc. from certain counties mainly which are mainly located in rural areas. The data from Google searches related to COVID-19 vaccination includes artificial noise. This noise is intentionally induced by Google to preserve users' privacy \citep{google_vsi_description}. Some of the datasets  changed the methods used for data collection during our period of study in an effort to improve their quality.  \par ($ii$)  \emph{Features used}: Our proposed model does not consider every possible feature that impacts VH behavior. We  only consider features for which there is publicly available data. \par 

{\bf Future Research Directions:} The scope of the proposed model can be extended as follows.\par
($i$) \emph{Vaccine supply chain}: Predictions about VH behavior can inform decisions about managing  the vaccine supply chain. VH impacts the demand, which in turn impacts the distribution of vaccines. Information about VH was particularly important in the early stages of the pandemic when there was a limited amount of vaccine available. One could use our proposed model to generate the data needed for models that support decisions related to vaccine distribution.  \par  
($ii$) \emph{VH for other vaccines}: Similar models can be developed to evaluate dynamic changes of VH for other vaccines. These models could shed light on the relationship between VH for COVID-19 and other vaccines. This information can be helpful in designing strategies to combat VH overall.  \par 
($iii$) \emph{Strategies to reduce VH}: There is a need for studies that can help us understand the impact of vaccination-related policies and COVID-19 restrictions on VH. Although our work shed some light on the impact of these strategies to reduce VH, more can be done.

\newpage

\bibliographystyle{informs2014}
\bibliography{ref}

\end{document}